\documentclass[12pt,epsf]{article}
\usepackage{graphicx}
\begin{document}
\def\a{\alpha}
\def\b{\beta}
\def\c{\varepsilon}
\def\d{\delta}
\def\e{\epsilon}
\def\f{\phi}
\def\g{\gamma}
\def\h{\theta}
\def\k{\kappa}
\def\l{\lambda}
\def\m{\mu}
\def\n{\nu}
\def\p{\psi}
\def\q{\partial}
\def\r{\rho}
\def\s{\sigma}
\def\t{\tau}
\def\u{\upsilon}
\def\v{\varphi}
\def\w{\omega}
\def\x{\xi}
\def\y{\eta}
\def\z{\zeta}
\def\D{\Delta}
\def\G{\Gamma}
\def\H{\Theta}
\def\L{\Lambda}
\def\F{\Phi}
\def\P{\Psi}
\def\S{\Sigma}
\def\o{\over}
\def\beq{\begin{eqnarray}}
\def\eeq{\end{eqnarray}}
\newcommand{\gsim}{ \mathop{}_{\textstyle \sim}^{\textstyle >} }
\newcommand{\lsim}{ \mathop{}_{\textstyle \sim}^{\textstyle <} }
\newcommand{\vev}[1]{ \left\langle {#1} \right\rangle }
\newcommand{\bra}[1]{ \langle {#1} | }
\newcommand{\ket}[1]{ | {#1} \rangle }
\newcommand{\EV}{ {\rm eV} }
\newcommand{\KEV}{ {\rm keV} }
\newcommand{\MEV}{ {\rm MeV} }
\newcommand{\GEV}{ {\rm GeV} }
\newcommand{\TEV}{ {\rm TeV} }
\def\diag{\mathop{\rm diag}\nolimits}
\def\Spin{\mathop{\rm Spin}}
\def\SO{\mathop{\rm SO}}
\def\O{\mathop{\rm O}}
\def\SU{\mathop{\rm SU}}
\def\U{\mathop{\rm U}}
\def\Sp{\mathop{\rm Sp}}
\def\SL{\mathop{\rm SL}}
\def\tr{\mathop{\rm tr}}
\def\IJMP{Int.~J.~Mod.~Phys. }
\def\MPL{Mod.~Phys.~Lett. }
\def\NP{Nucl.~Phys. }
\def\PL{Phys.~Lett. }
\def\PR{Phys.~Rev. }
\def\PRL{Phys.~Rev.~Lett. }
\def\PTP{Prog.~Theor.~Phys. }
\def\ZP{Z.~Phys. }
\baselineskip 0.7cm
\begin{titlepage}
\begin{flushright}
UT-04-14
\end{flushright}

\vskip 1.35cm
\begin{center}
{\large \bf
A solution to the baryon and dark-matter coincidence puzzle in a
  $\tilde{N}$ dominated early universe
}
\vskip 1.2cm
M.~Ibe${}^{1}$ and T.~Yanagida${}^{1,2}$
\vskip 0.4cm

${}^1${\it Department of Physics, University of Tokyo,\\
     Tokyo 113-0033, Japan}

${}^2${\it Research Center for the Early Universe, University of Tokyo,\\
     Tokyo 113-0033, Japan}

\vskip 1.5cm

\end{center}

\abstract{
If a bosonic partner of a right-handed neutrino dominates the early
 universe sufficiently before its decay, important ingredients in the
 present universe are related to physics of the right-handed neutrino 
 sector. 
In particular, we find that the ratio of the baryon to the dark-matter
 densities is given only by low-energy parameters such as a neutrino
 mass and a gravitino mass if the reheating temperature of inflation
 is much higher than $10^{12}$~GeV.
Here, the gravitino is assumed to be the lightest  supersymmetric
 particle and the dominant component of the dark matter.
The observed  ratio,  $\Omega_{B}/\Omega_{DM} \simeq 0.21\pm 0.04$,
 suggests the mass of the gravitino to be in the range of $\cal O$(10)~MeV  
 provided the CP violating phase is of the order 1.   
 }
\end{titlepage}
\setcounter{page}{2}

\section{Introduction}
The seesaw mechanism~\cite{seesaw} is very attractive, since it explains
naturally not only the observed small neutrino masses but also the
baryon asymmetry in the present universe~\cite{LGorg}.
The important ingredient in the seesaw mechanism is the presence of
right-handed neutrinos $N_i (i = 1 - 3)$ whose Majorana masses $M_i$ are
very large such as $M_i\simeq 10^{9-15}$~GeV.
In a supersymmetric (SUSY) extension of the seesaw mechanism the
right-handed neutrinos $N_i$ are necessarily accompanied with
SUSY-partner bosons $\tilde{N}_i$ (right-handed sneutrinos), and it is
quite plausible~\cite{MY} that the $\tilde{N_i}$ have very large
classical values during inflation if the masses of the right-handed
neutrinos are smaller than the Hubble constant of the inflation.
If it is the case and coherent oscillations of the bosons $\tilde{N}_i$
dominate the early universe sufficiently before their decays, some of
important parameters in the present universe are determined by the
physics of the right-handed neutrino sector. 
In this letter, we point out that if a boson partner of a right-handed
neutrino $N_1$ dominate once the early universe it may solve the
coincidence puzzle of the baryon and dark-matter densities provided that
the mass of gravitino is ${\cal O}(10)$~MeV.

Before discussing the physics of $\tilde{N}_1$ we should note a generic
problem in supergravity, that is the gravitino problem~\cite{gravitino}.  
If the gravitino is unstable, it has a long lifetime and decays during
or after the big-bang nucleosynthesis (BBN). 
The decay products destroy the light elements created by the BBN and
hence the abundance of the relic gravitino is constrained from above.   
This leads to an upper bound of the reheating temperature $T_R$ of
inflation.  
The recent detailed analysis~\cite{KKM} shows a stringent upper bound
such as $T_R<10^4$ GeV for the gravitino having hadronic decay modes.   
In the present scenario this reheating temperature means the
temperature just after the decay of the coherent $\tilde{N}$ oscillation
(i.e. the decay temperature $T_d$). 
Such a low decay temperature is nothing unnatural in the scenario, but
the produced lepton (baryon) asymmetry is too small~\cite{MY}. 

A solution to this gravitino problem is to assume that the gravitino is
the lightest SUSY particle (LSP) and hence stable~\cite{BBP}.
This solution is very interesting in the present scenario, since the
ratio of $\Omega_{B}$ to $\Omega_{3/2}$ is independent of the unknown
temperature $T_d$, but it is given by only low-energy parameters if the
reheating temperature of inflation is sufficiently high. 
Here, $\Omega_B$ and $\Omega_{3/2}$ are mass density parameters of the
baryon and the gravitino, respectively.
We find that the ratio is determined by masses of a neutrino, the gluino
and the gravitino and an effective CP-violating phase (as shown in
Eq.~(\ref{eq:ratio})). 
The observation $\Omega_{B}/\Omega_{DM} \simeq 0.21\pm 0.04$~\cite{WMAP}
suggests $m_{3/2} = {\cal O}$(10)~MeV. ($m_{3/2}$ is the mass of the
gravitino.) 
Here, we have assumed that the gravitino is the dominant component of
the cold dark matter, that is $\Omega_{\rm DM}\simeq \Omega_{3/2}$.
The gravitino of mass in the range of ${\cal O}$(10)~MeV will be testable
in future experiments as discussed in Ref.~\cite{BHRY}.

\section{Matter from a coherent right-handed  sneutrino} 

\subsection{Baryon asymmetry from a coherent right-handed sneutrino} 
We consider a frame work of the minimal supersymmetric standard model
(MSSM) with three generations of heavy right-handed neutrinos $N_i$
$(i=1-3)$.  
The $N_i$ couple to the MSSM particles through a superpotential,
\begin{eqnarray}
 W=\frac{1}{2}M_i N_i N_i + h_{i \alpha} L_{\alpha}H_{u}N_i,
\label{eq:superpotential1}
\end{eqnarray}
where $M_i$ denote masses of the right-handed neutrinos and
$L_{\alpha}$ $(\alpha= e,\mu,\tau)$ and $H_u$ are the supermultiplets of
lepton doublets and a Higgs doublet which couples to up-type
quarks.
The small left-handed neutrino masses are obtained via the seesaw
mechanism~\cite{seesaw}.

The right-handed sneutrinos may have large classical values
during inflation if their effective masses are smaller than the Hubble
parameter $H_{\rm inf}$ \cite{MY}.
Hereafter, we restrict our discussion to the lightest right-handed
sneutrino $\tilde{N}_1$, for simplicity, and treat the amplitude
$\tilde{N}_{1}^{\rm init}$ during the inflation as a free parameter.

After the end of the inflation, the Hubble parameter $H$ decreases and
the $\tilde{N}_1$ starts to oscillate when $H$ becomes smaller than its mass  
$M_1$.\footnote{  
We assume the potential for the $\tilde{N}_1$ is given by a mass term, $V =
M_1^2 |\tilde{N}_1|^2$. 
We discuss the validity of this simplification of the potential for
analyzing the dynamics of the $\tilde{N}_1$ in the next section. 
} 
The coherent oscillation of the $\tilde{N}_1$ decays into $L\tilde{H}_u$ or
$\tilde{L}H_u$ and their CP-conjugates when $H\simeq\Gamma_{N_1}$, where
$\Gamma_{N_1} \simeq (1/4\pi)\sum_\alpha |h_{1\alpha}|^2 M_1$ is the
decay rate of the $\tilde{N}_1$.
The decay produces the lepton number density as, $n_L=\epsilon \times
n_{\tilde{N}_1}$, where $n_{\tilde{N}_1}$ is the number density of 
the $\tilde{N}_1$ at the decay time, and $\epsilon$ is the lepton asymmetry
produced in the $\tilde{N}_1$ decay.   
Assuming $M_1\ll M_2,M_3$, the explicit form of $\epsilon$ is given
by~\cite{BY,ep-calc} 
\begin{eqnarray}
  \label{eq:CPasym}
 \epsilon
 \simeq (1 - 2)\times 10^{-10} 
  \bigg(\frac{M_1}{10^6\GEV}\bigg)
  \bigg(\frac{m_{\nu 3}}{0.05 {\rm eV}}\bigg)\sin\delta_{\rm eff},
\end{eqnarray} 
where $\delta_{\rm eff}$ is an effective CP violating phase and $m_{\nu
3}$ corresponds to the heaviest neutrino mass, we have used
$\vev{H_u}=174\, \GEV\times\, \sin\beta$, assuming $\sin\beta\simeq
1/\sqrt{2} - 1$. 

When the $\tilde{N}_1$ dominates the universe, we can write the energy
density ($\rho$) and the entropy density ($s$) of the universe at the
decay time as 
\begin{eqnarray}
\label{eq:energyconsv}
\rho &\simeq& M_1^2 |\tilde{N}_{1}^{\rm decay}|^2 
     \simeq \frac{\pi^2}{30}g_*(T_d) T_{d}^4
     \simeq 3M_{\rm pl}^2 \Gamma_{N_1}^2,\\ \nonumber
s &\simeq& \frac{2\pi^2}{45}g_*(T_d) T_{d}^3.
\end{eqnarray}
Here, $T_d$ is the temperature of radiation right after the
$\tilde{N}_1$ decay, $g_*$ the number of effective degrees of
freedom which is 230 for the temperature $T\gg 1$ TeV in the
MSSM and $M_{\rm pl}\simeq 2.4\times 10^{18}$~GeV the reduced Planck scale.  
In the above equation, we have assumed instantaneous decay of
the $\tilde{N}_1$ and used the energy conservation.
Hereafter, we only focus on the scenario in which the $\tilde{N}_1$
domination is the case. 

The resultant lepton number is converted to the baryon-number
asymmetry~\cite{LGorg}, which is given by~\cite{MY,Hamaguchi:2001gw}   
\begin{eqnarray}
\label{eq:Baryonasymm}
 \frac{n_B}{n_{\gamma}}
&=& -\frac{8}{23}\bigg(\frac{n_L}{n_{\gamma}}\bigg)
= -\frac{8}{23}\bigg(\frac{n_L}{\rho}\bigg)\bigg(\frac{\rho}{s}\bigg)
\bigg(\frac{s}{n_{\gamma}}\bigg)
= -\frac{8}{23}\bigg(\frac{\epsilon}{M_1}\bigg)\bigg(\frac{3T_d}{4}\bigg)
\bigg(\frac{s}{n_{\gamma}}\bigg)\\\nonumber
&\simeq &
(1.7 - 3.4) \times 10^{-10}
\bigg(\frac{T_d}{10^6~{\rm GeV}}\bigg)
\bigg(\frac{m_{\nu 3}}{0.05~{\rm eV}}\bigg)
\sin\delta_{\rm eff},
\end{eqnarray}
where we have used $s/n_{\gamma}\simeq 7.04$ at the present and
Eqs.~(\ref{eq:energyconsv}) in the last equation.
We take $m_{\nu 3}\simeq 0.05$ eV as suggested from the atmospheric
neutrino oscillation and assume $\sin\delta_{\rm eff}\simeq 1$.       
Then, the observed baryon asymmetry
$n_B/n_{\gamma}=6.5^{+0.4}_{-0.3}\times 10^{-10}$~\cite{WMAP} implies 
\begin{eqnarray}
T_d\simeq 10^6~{\rm GeV}-10^7~{\rm GeV}.
\label{eq:decaytemp}
\end{eqnarray}

Before closing this subsection, we should mention washout effects of
the lepton asymmetry.
When the decay temperature of the $\tilde{N}_1$ is close to its
mass, $T_d\simeq M_1$, the produced lepton-number asymmetry is washed
out by lepton-number violating interactions mediated by $N_1$.
Thus, in order to avoid the washout effect, we require $T_d<M_1$, and
this condition is rewritten by using Yukawa coupling constants in
Eq.~(\ref{eq:superpotential1}) as \cite{Hamaguchi:2001gw}, 
\begin{eqnarray}
\label{eq:washout}
\bigg(\sum_{\alpha}|h_{1\alpha}|^2\bigg)^{\frac{1}{2}}
\simeq 5\times 10^{-6}
\bigg(\frac{T_d}{10^6~{\rm GeV}}\bigg)^{\frac{1}{2}}
\bigg(\frac{T_d}{M_1}\bigg)^{\frac{1}{2}}
<
5\times 10^{-6} 
\bigg(\frac{T_d}{10^6~{\rm GeV}}\bigg)^{\frac{1}{2}}.
\end{eqnarray}
Here, we have used Eq.~(\ref{eq:energyconsv}) to relate
$M_1$ and $T_d$.
We require the Yukawa couplings $h_{1\alpha}$ to be as small as
the Higgs coupling to the electron.
We may explain naturally such small Yukawa coupling constants by a
spontaneously broken discrete $Z_6$ flavor symmetry
\cite{Hamaguchi:2001gw,Fujii:2001zr}.

\subsection{Conditions for $\tilde{N}$ domination}
In the previous subsection, we consider the $\tilde{N}_1$ to dominate
the energy density of the early universe.  
We discuss, here, conditions for the $\tilde{N}$ domination.

We classify the history of the energy density of the early universe by
the reheating temperature $T_R$ of inflation, the initial amplitude 
$|\tilde{N}_1^{\rm  init}|$ and the decay temperature $T_d$ of the
right-handed sneutrino.\footnote{ 
$T_R$ is defined as a temperature of the radiation right after the end
of the reheating process of inflation.
}
If the Hubble parameter at the end of the reheating process of inflation
is smaller than $M_1$, the $\tilde{N}_1$ starts to oscillate around its
minimum before the end of the reheating.
The domination of the $\tilde{N}_1$ starts at a temperature $T_{\rm dom}$
which is estimated as
\begin{eqnarray}
\label{eq:Tdom}
 T_{\rm dom}\simeq
T_R\times \bigg(\frac{|\tilde{N}_1^{\rm init}|^2}{3M_{\rm pl}^2}\bigg).
\end{eqnarray}
Thus, a condition for the domination of the $\tilde{N}_1$ is 
\begin{eqnarray}
\label{eq:dominate1}
 T_{\rm dom} \simeq T_R\times 
\bigg(\frac{|\tilde{N}_1^{\rm init}|^2}{3M_{\rm pl}^2}\bigg)
> T_d.
\end{eqnarray}

On the other hand, if the Hubble parameter at the end of the reheating
of inflation is larger than $M_1$, the $\tilde{N}_1$ starts to oscillate
after the end of the reheating process. 
The temperature of the background radiation when the $\tilde{N}_1$
oscillation starts is given by
\begin{eqnarray}
\label{eq:oscillate}
 T_{\rm osci}\simeq \bigg(\frac{90}{\pi^2 g_*}\bigg)^{1/4}\sqrt{M_{\rm pl} M_1}. 
\end{eqnarray}
In this case, the $\tilde{N}_1$ dominates the universe soon after it
starts the oscillation.
As in the previous case the temperature $T_{\rm dom}$ at which the
domination of the $\tilde{N}_1$ begins is estimated as 
\begin{eqnarray}
\label{eq:Tdom2}
 T_{\rm dom} \simeq
T_{\rm osci}\times \bigg(\frac{|\tilde{N}_1^{\rm init}|^2}{3M_{\rm pl}^2}\bigg).
\end{eqnarray}
Thus, the condition for the $\tilde{N}_1$ to dominate the
universe is,
\begin{eqnarray}
\label{eq:dominate2}
 T_{\rm dom} &\simeq& 
T_{\rm osci}\times\bigg(\frac{|\tilde{N}_1^{\rm init}|^2}{3M_{\rm pl}^2}\bigg)
> T_d.
\end{eqnarray}
As we have seen in the previous subsection, we consider $T_d\simeq
10^6~\GEV-10^7$~GeV, and hence the above conditions Eqs.~(\ref{eq:dominate1})
or (\ref{eq:dominate2}) can be satisfied for a wide range of the initial
amplitude of the $\tilde{N}_1$, $T_R$ and $M_1$.  

\subsection{Dark-matter genesis}
As discussed in the introduction, we assume the gravitino to be the
LSP and the dominant component of the cold dark-matter (CDM).   
As we see below, the relic gravitino density is proportional to the
decay temperature of the $\tilde{N}_1$ if the gravitinos are produced
mainly by the $\tilde{N}_1$ decay. 
Thus, the ratio between $\Omega_{B}h^2$ and $\Omega_{\rm 3/2}h^2$
becomes independent of the decay temperature $T_d$ (see
Eq.~(\ref{eq:Baryonasymm})) and is determined only by low-energy
parameters.   

However, the gravitino is forced into thermal equilibrium by the
scattering process if the decay temperature $T_d$ is sufficiently high.   
If it is the case, the density of the gravitino is not proportional
to $T_d$, making the above argument invalid. 
The freeze-out temperature of the gravitino from the thermal bath is
given by~\cite{Fujii:2002fv}
\begin{eqnarray}
\label{eq:freezout}
T_f \simeq 10^9~{\rm GeV}
\bigg(\frac{g_*(T_f)}{230}\bigg)^{1/2}
\bigg(\frac{m_{3/2}}{10~{\rm MeV}}\bigg)^2
\bigg(\frac{1 \rm TeV}{m_{\rm gluino}}\bigg)^2,
\end{eqnarray}
where $m_{\rm gluino}$ denotes the mass of the gluino.
We should note here that our conclusion does not change as long as $T_d
< T_f$.
We check in the next subsection that this condition is satisfied. 

On the other hand, when the reheating temperature $T_R$ of inflation is
higher than $T_f$, the gravitino is kept in the thermal equilibrium and
its resultant density is estimated as
\begin{eqnarray}
\label{eq:thermalrelic}
 \Omega_{3/2}h^2 \simeq 5.0 \times 10^3\,
\bigg(\frac{m_{3/2}}{10~{\rm MeV}}\bigg)\,
\bigg(\frac{230}{g_*(T_f)}\bigg).
\end{eqnarray}
If $T_R$ is lower than $T_f$, the gravitino cannot be in the thermal
equilibrium and its resultant density is given by~\cite{Bolz:2000fu} 
\begin{eqnarray}
\label{eq:freezedrelic}
 \Omega_{3/2}h^2 \simeq 2.1 \times 10^3\,
  \bigg(\frac{T_R}{10^{10}~{\rm GeV}}\bigg)\,
  \bigg(\frac{10~{\rm MeV}}{m_{3/2}}\bigg)\,
  \bigg(\frac{m_{\rm gluino}}{1{\rm TeV}}\bigg)^2.
\end{eqnarray}

However, the gravitino density from the reheating process of the
inflation is diluted by entropy production from the $\tilde{N}_1$ decay.   
By assuming the instantaneous decays of the $\tilde{N}_1$, which is
accurate enough for the present purpose, we obtain the dilution
factor (from the energy conservation) as
\begin{eqnarray}
  \label{eq:dilution}
 \Delta\equiv\bigg(\frac{s_{\rm after}}{s_{\rm before}}\bigg) 
 \simeq\frac{T_{\rm dom}}{T_d} \simeq
  \left\{
   \begin{array}{lll}
   &\displaystyle{\frac{T_R}{T_d}}
     \bigg(\frac{|\tilde{N}_1^{\rm init}|^2}{3M_{\rm pl}^2}\bigg)& 
     \,\,(T_R<T_{\rm osci}),\\
   & \displaystyle{\frac{T_{\rm osci}}{T_d}}
     \bigg(\frac{|\tilde{N}_1^{\rm init}|^2}{3M_{\rm pl}^2}\bigg)&
     \,\,(T_R>T_{\rm osci}),\\
   \end{array}
  \right.
\end{eqnarray}
where we have used Eq.~(\ref{eq:dominate1}) and (\ref{eq:dominate2}).
As a result, the present gravitino density is written as 
\begin{eqnarray}
\label{eq:gravitinorelic}
 \Omega_{3/2}h^2 = 
\Omega_{3/2}(T_d)h^2 + \frac{1}{\Delta}\Omega_{3/2}(T_R)h^2,
\end{eqnarray}
where $\Omega_{3/2}(T)h^2$ denote the gravitino density in
Eqs.~(\ref{eq:thermalrelic}) or (\ref{eq:freezedrelic}) at each
temperatures $T = T_d$ or $T_R$.
The first term in Eq.~(\ref{eq:gravitinorelic}) represents the density
of the gravitino produced in the $\tilde{N}_1$ decay, while the second
term is the resultant density of the gravitino produced in the reheating
process of inflation.  

\subsection{A solution to the coincidence puzzle}
As we have seen, the baryon asymmetry in the present universe comes from
the $\tilde{N}_1$ decay, and the resultant baryon density
$\Omega_{B}h^2$ is given by 
\begin{eqnarray}
 \label{eq:Baryon}
  \Omega_Bh^2\simeq (6.3-13) \times 10^{-3} 
  \bigg(\frac{T_d}{10^6~{\rm GeV}}\bigg)
  \bigg(\frac{m_{\nu 3}}{0.05~{\rm eV}}\bigg)
  \sin\delta_{\rm eff},
\end{eqnarray}
where we have used the proton mass $m_p\simeq 0.938$~GeV. (See
Eq.~(\ref{eq:Baryonasymm}).) 
On the other hand, from Eq.~(\ref{eq:gravitinorelic}) the dark
matter (the gravitino LSP) density is written as 
\begin{eqnarray}
 \label{eq:DM}
  \Omega_{\rm DM}h^2\simeq 0.21 \times 
  \bigg(\frac{T_d}{10^{6}~{\rm GeV}}\bigg)\,
  \bigg(\frac{10~{\rm MeV}}{m_{3/2}}\bigg)\,
  \bigg(\frac{m_{\rm gluino}}{1{\rm TeV}}\bigg)^2\times k_N,
\end{eqnarray}
where $k_N$ is defined as
\begin{eqnarray}
k_N = \left\{
   \begin{array}{rcll}
    \label{eq:gravitinofrac} 
     1&+&0.2\times 
     \displaystyle{\bigg(\frac{10^{12}~{\rm GeV}}{T_R}\bigg)
     \bigg(\frac{m_{3/2}}{30~\MEV}\bigg)^2
     \bigg(\frac{1{\rm TeV}}{m_{\rm gluino}}\bigg)^2
     \bigg(\frac{3M_{\rm pl}^2}{|N_1^{\rm init}|^2}\bigg)}
     & (T_R>T_f, \,\,T_R<T_{\rm osci}),\\
    1&+&0.2\times 
     \displaystyle{\bigg(\frac{10^{12}~{\rm GeV}}{T_{\rm osci}}\bigg)
     \bigg(\frac{m_{3/2}}{30~\MEV}\bigg)^2
     \bigg(\frac{1{\rm TeV}}{m_{\rm gluino}}\bigg)^2
     \bigg(\frac{3M_{\rm pl}^2}{|N_1^{\rm init}|^2}\bigg)}
     & (T_R>T_f, \,\, T_R>T_{\rm osci}),\\
    1&+&\displaystyle{\bigg(\frac{3M_{\rm pl}^2}{|N_1^{\rm init}|^2}\bigg)}
     & (T_R<T_f, \,\, T_R<T_{\rm osci}),\\
    1&+&\displaystyle{\bigg(\frac{T_R}{T_{\rm osci}}\bigg)
     \bigg(\frac{3M_{\rm pl}^2}{|N_1^{\rm init}|^2}\bigg)}
     & (T_R<T_f,\,\, T_R>T_{\rm osci}),
   \end{array}
  \right.\nonumber\\
\end{eqnarray}
for each values of $T_R$, $T_f$ and $T_{\rm osci}$.

For the third and the fourth cases in Eq.~(\ref{eq:gravitinofrac}), the
gravitino densities depend on the initial amplitudes of the
$\tilde{N}_1$. 
On the other hand,  the second terms are negligible for the first and
the second cases in  Eq.~(\ref{eq:gravitinofrac}) if the reheating
temperature $T_R$ or the  oscillation temperature $T_{\rm osi}$ are much 
higher than $10^{12}$~GeV. 
(The model discussed in the next section gives most likely
$|\tilde{N}_1^{\rm init}|\simeq M_{\rm pl}$.)

\begin{figure}
 \begin{center}
  \begin{minipage}{0.49\linewidth}
   \begin{center}
    \includegraphics[width=.95\linewidth]{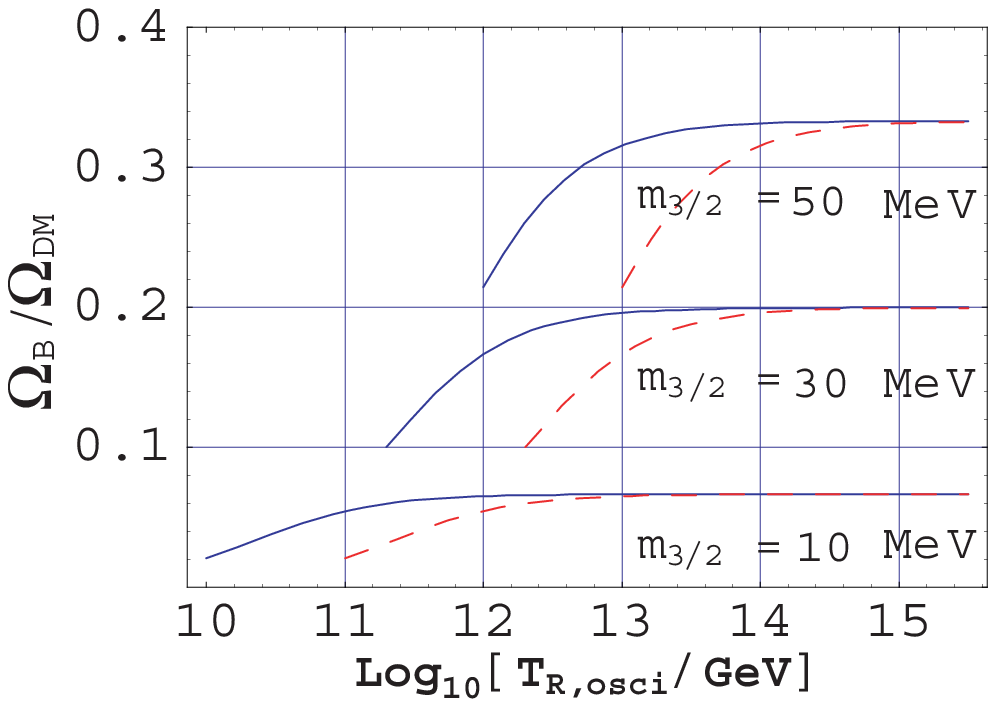}
   \end{center}
  \end{minipage}
  \begin{minipage}{0.49\linewidth}
   \begin{center}
    \includegraphics[width=.95\linewidth]{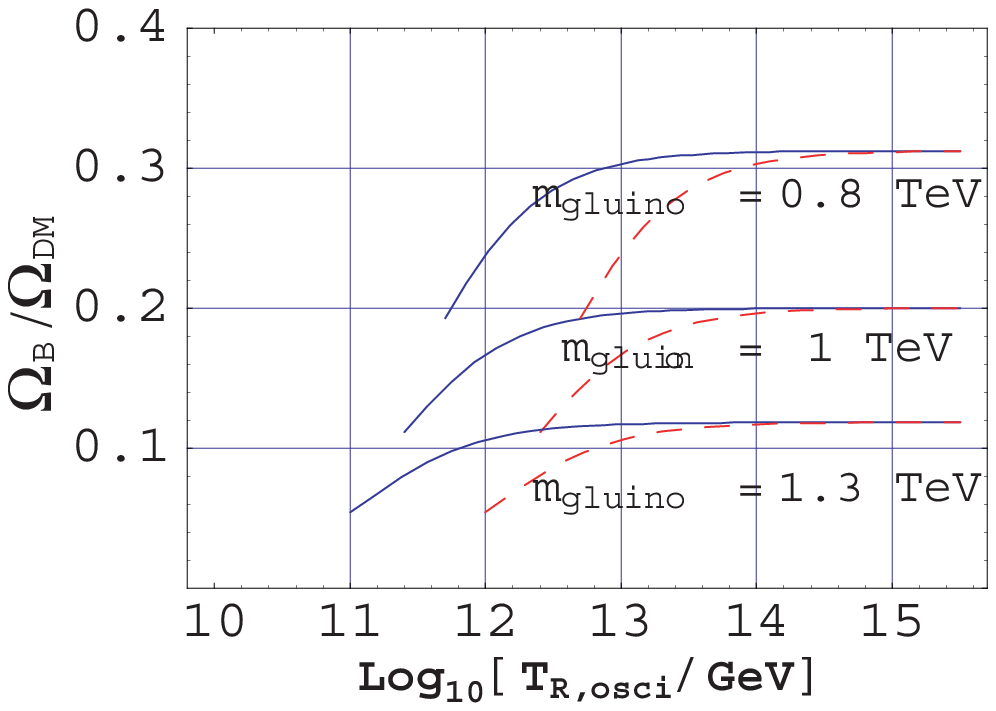}
   \end{center}
  \end{minipage}
 \end{center}
 \caption{The ratio $\Omega_{B}/\Omega_{\rm DM}$ as a function of the
 $T_{R,{\rm osi}}$ (see Eq.~(\ref{eq:gravitinofrac})).
In the left panel, solid (dashed) lines correspond to the ratio for
 $m_{3/2}=10$~MeV, 30~MeV, and 50~MeV from the bottom up with 
 $m_{\rm gluino}=1$~TeV and  $|\tilde{N}_1^{\rm init}|^2/3 M_{\rm pl}^2=1$ ($|\tilde{N}_1^{\rm init}|^2/3 M_{\rm pl}^2=0.1$).
In the right panel, solid (dashed) lines correspond to the ratio for
 $m_{\rm gluino}=0.8$~TeV, 1~TeV, and 1.3~TeV from the bottom up with 
 $m_{3/2}=30$~MeV and  $|\tilde{N}_1^{\rm init}|^2/3 M_{\rm pl}^2=1$
 ($|\tilde{N}_1^{\rm init}|^2/3 M_{\rm pl}^2=0.1$).  
In this calculation, we have fixed $m_{\nu 3}= 0.05$~eV,
 $\sin\delta_{\rm eff}=1$, and $\tan\beta =1$.
 }
 \label{fig:ratio}
\end{figure}

In Fig.~\ref{fig:ratio}, we plot the ratio $\Omega_{B}/\Omega_{\rm DM}$
as a function of $T_{R,{\rm osci}}$ for the first and the second cases
in Eq.~(\ref{eq:gravitinofrac}).
We find that the ratio becomes independent of the $|\tilde{N}_1^{\rm
init}|$ and $T_{R,{\rm osci}}$ for sufficiently high temperatures
$T_{R,{\rm osci}}$ and it is determined only by the low-energy
parameters.  
In those regions, the ratios $\Omega_{B}/\Omega_{\rm DM}$ are given by  
\begin{eqnarray}
 \label{eq:ratio}
  \frac{\Omega_Bh^2}{\Omega_{\rm DM}h^2}\simeq (0.1-0.2)
  \bigg(\frac{m_{3/2}}{30~{\rm MeV}}\bigg)\,
  \bigg(\frac{1{\rm TeV}}{m_{\rm gluino}}\bigg)^2
  \bigg(\frac{m_{\nu 3}}{0.05~{\rm eV}}\bigg)
  \sin\delta_{\rm eff}.
\end{eqnarray}
Comparing Eq.~(\ref{eq:ratio}) with the WMAP result
$\Omega_B/\Omega_{\rm DM}\simeq 0.21\pm 0.04$~\cite{WMAP}, we obtain
the mass of the gravitino as   
\begin{eqnarray}
 \label{eq:gravitinomass}
  m_{3/2} \simeq 30~{\rm MeV} - 60~{\rm MeV},  
  \,\,\,
  \mbox{for }
  \sin\delta_{\rm eff} \simeq 1,
  \,\,\,
  m_{\rm gluino} \simeq 1~{\rm TeV},
\end{eqnarray}
which suggests a gauge mediation SUSY breaking
(GMSB)~\cite{gauge-mediation}.  
Therefore, the coincidence puzzle between the baryon and the 
dark-matter densities can be naturally solved in the GMSB model when
the both of the densities come dominantly from the $\tilde{N}_1$ decay.
Notice that we obtain $T_f\simeq 10^{10}$~GeV from
Eq.~(\ref{eq:freezout}) in the parameter region 
Eq.~(\ref{eq:gravitinomass}) and hence the condition $T_d<T_f$ discussed  
in the previous subsection is satisfied since $T_d\simeq 10^{6-7}$~GeV.

Finally, we comment on constraints from the Big Bang Nucleosynsesis
(BBN).
For the gravitino LSP scenario, the next to the lightest SUSY
particle (NLSP) has a long lifetime and it may spoil the success of the
BBN, in general.
However, in our scenario of $m_{3/2}= {\cal O}(10)$~MeV, the lifetime of
the NLSP is sufficiently short as 
\begin{eqnarray}
 \label{eq:lifetime}
  \tau_{\rm NLSP} 
  \simeq 2\times 10^{-2} {\rm sec.} \bigg(\frac{m_{3/2}}{10~{\rm MeV}}\bigg)^2
  \bigg(\frac{300~{\rm GeV}}{m_{\rm NLSP}}\bigg)^5.
\end{eqnarray}
Thus, the NLSP can escape from the BBN constraints~\cite{Asaka:2000zh}.

\section{Some discussion}
\subsection{A model for the right-handed neutrino sector}

In the previous section, we have used a potential for the
$\tilde{N}_1$, $V\simeq M_1^2 |\tilde{N}_1|^2$.  
However, if we assume the broken $U(1)_{B-L}$ gauge symmetry to generate
the Majorana masses of $N_i$, the other MSSM fields are destabilized
through a D-term potential of $U(1)_{B-L}$ during the  $\tilde{N}_1$
oscillation.\footnote{
This problem is not present, if the gauge coupling constant of
$U(1)_{B-L}$ is extremely small.   
} 
In this case, the evolution of the scalar fields becomes rather complex 
to trace and hence it becomes difficult to predict the cosmic baryon 
asymmetry.\footnote{ 
If $B$ or $L$ violating non-renormalizable terms exist in the MSSM
superpotential, the Affleck-Dine baryogenesis \cite{ADBaryo} may work,
which changes our result in the previous section. 
Even if there is no such $B$ or $L$ violating terms, decay processes of
the multi-field oscillations are not so simple and the fate of the
$\tilde{N}_1$ oscillation is difficult to be predicted.   
}

To avoid the above problem, we consider a model with $U(1)_R\times
Z_4^{B-L}$ symmetry whose charge assignments are given in
Table~\ref{tab:charge}.\footnote{ 
The three right-handed neutrinos are required to cancel $Z_4^{B-L}$ gauge
anomalies.}
Here, $U(1)_R$ is the $R$ symmetry.
A simple superpotential allowed by the symmetry is 
\begin{eqnarray}
\label{eq:superpotential2}
 W = y X (S^2-v^2)-\frac{1}{2}f_i S N_i^2 + h_{i \alpha} L_{\alpha}H_{u}N_i 
+ W_{\rm MSSM},
\end{eqnarray}
where we have added two MSSM singlets $X$ and $S$, $y$ and $f_i$ denote
the Yukawa coupling constants, the parameter $v$ the breaking scale of
the $Z^{B-L}_4$ symmetry, and $W_{\rm MSSM}$ the superpotential consists
of the MSSM fields. 
As we see below, the evolution of the $\tilde{N}_1$ can be analyzed by  
using the potential $M_1^2|\tilde{N}_1|^2$ as long as  the Hubble
parameter during the inflation is much smaller than $v \simeq
10^{15}$~GeV.

From the superpotential Eq.~(\ref{eq:superpotential2}), the scalar
potential which is relevant to the dynamics of the $\tilde{N}_1$ is
given by  
\begin{eqnarray}
\label{eq:scalarpotential}
 V =  |y(\tilde{S}^2-v^2)|^2 + |2 y \tilde{S}\tilde{X} 
  - f \frac{1}{2}\tilde{N}_1^2|^2 
  + |h\phi^2-f \tilde{S}\tilde{N}_1|^2 + |h \tilde{N}_1 \phi|^2,
\end{eqnarray}
where $\phi$ denotes the flat direction in the MSSM defined by $H_u =
1/\sqrt{2} (0,\phi)^T$, $\tilde{L} = 1/\sqrt{2} (\phi,0)^T$, and we
have omitted the flavor index from the Yukawa coupling constants for
abbreviation.\footnote{
We can easily extend our discussion to the case where the Hubble mass
terms are induced by the supergravity effects. 
}
In the following discussion, we focus on the evolution of the
$\tilde{N}_1$, $\tilde{X}$ and $\tilde{S}$, assuming $M_1\ll H_{\rm
inf}\ll M_2,M_3$ and $\phi=0$.    
The dynamics of $\phi$ is discussed in the next subsection, where we see
that the thermal mass term sets $\phi$ to the origin.

If $H_{\rm inf}\ll v$, $\tilde{S}$ and $\tilde{X}$ are fixed to
their minima during inflation (we have required $y$ be not too small).
We also require  $f|\tilde{N}_1^{\rm init}| \ll v$ not to destabilize
the minimum of $\tilde{S}$.\footnote{ 
Even for $|\tilde{N}_1^{\rm init}|\simeq M_{\rm pl}$ this condition
can be easily realized by a spontaneously broken discrete $Z_6$
symmetry~\cite{Hamaguchi:2001gw,Fujii:2001zr}, where $f$ may be as
small as $10^{-5}$. 
}
Thus, the scalar fields are fixed in the end of inflation at 
\begin{eqnarray}
\label{eq:minima}
 \tilde{N}_1 =\tilde{N}_1^{\rm init},\,\,\,\,
  \tilde{X} = \bigg(\frac{f}{4y v}\bigg)(\tilde{N}_1^{\rm init})^2,\,\,\,\,
  \tilde{S} = v.
\end{eqnarray}

After the end of inflation, the Hubble parameter $H$ becomes smaller than
$M_1$ and the $\tilde{N}_1$ starts to oscillate around its origin.  
Since the time scale of the motion of  $\tilde{X}$ and $\tilde{S}$  
($\sim 1/(yv)$) is much smaller than the one of the $\tilde{N}_1$
oscillation ($\sim 1/M_1$),  $\tilde{X}$ and $\tilde{S}$ trace their
minima along with the $\tilde{N}_1$ oscillation;
\begin{eqnarray}
\label{eq:motion1}
 \tilde{X}(t)\simeq \bigg(\frac{f}{4 y v}\bigg)\tilde{N}_1(t)^2,\,\,\,\,
  \tilde{S}(t)\simeq v.
\end{eqnarray}
Therefore, we find that our assumption in the previous sections to take
the scalar potential of the $\tilde{N}_1$ as $V\simeq M_1^2
|\tilde{N}_1|^2$ is valid.
Thus, we expect the initial amplitude of the $\tilde{N}_1^{\rm init}$ to
be of the order of $M_{\rm pl}$.\footnote{
Our approximation of the potential Eq.~(\ref{eq:scalarpotential}) is no
longer valid for $\tilde{N}_1\gg M_{\rm pl}$ in the supergravity
theory. 
}

\begin{table}[htb]
\begin{center}
\begin{tabular}{|c||c|c|c|c|c|c|}
\hline
Fields    & $Q_L$, $\bar{U}_R$, $\bar{E}_R$ & $L_L$, $\bar{D}_R$ 
&$H_u$, $H_d$ &$N_i$ & $X$ & $S$\\
\hline
$R$ charges & 1 & 1 & 0 & 1 & 2 &0 \\
\hline
$Z_4$ charges &1 & -3 & 2 & 1 & 0&2 \\
\hline
\end{tabular}
\caption{
Here, $Q_L$ and $L_L$ denote the SU(2)$_L$ doublet quarks and leptons,
 $\bar{U}_R$, $\bar{D}_R$ and $\bar{E}_R$ are the  SU(2)$_L$ singlet
 up- and down-quarks and leptons, and $H_{u,d}$ the up-type and 
 down-type Higgs.
}
\label{tab:charge}
\end{center}
\end{table}

\subsection{Stability of the $LH_u$ flat direction}
We give a comment on stability of the $LH_u$ flat direction $\phi$
during the $\tilde{N}_1$ oscillation. 
Instability of the $LH_u$ direction comes from a cross term in
the scalar potential between the $LH_u$ direction and the $\tilde{N}_1$ in
Eq.~(\ref{eq:scalarpotential}).\footnote{
We thank K. Hamaguchi for pointing out this problem.
} 
However, we find that thermal effects stabilize the $LH_u$
direction $\phi$
 
The $LH_u$ flat direction $\phi$ is at the origin when the $\tilde{N}_1$
has a large amplitude, since it has a large positive mass term
$|h\tilde{N}_1|^2|\phi |^2$.  
After the $\tilde{N}_1$ starts to oscillate, the positive mass term 
$|h\tilde{N}_1|^2|\phi|^2$ decreases and the cross term between $\phi$
and the $\tilde{N}_1$ becomes more significant than the positive mass term.
When the $\tilde{N}_1$ becomes smaller than $M_1/h$, (see the last two
terms in Eq.~(\ref{eq:scalarpotential})), $\phi$ seems to depart from
the origin for the $\tilde{N}_1\lsim M_1/h$. 
However, we should note here that there is a thermal mass term for $\phi$
from the thermal background, and hence the effective potential for
$\phi$ is given by 
\begin{eqnarray}
\label{eq:scalarpotential2}
 V\simeq  
|h\phi^2-M_1\tilde{N}_1|^2 + |h \tilde{N}_1 \phi|^2+ \alpha^2 T^2 |\phi|^2,
\end{eqnarray}
where $T$ denotes the temperature of the thermal background.
Here, the coefficient $\alpha$ is estimated as $\alpha^2 \simeq 3 g_2^2/8
+ g_1^2/8 \simeq 1/4$ for $\phi\ll T$, and we have omitted the thermal
effects for the $\tilde{N}_1$.\footnote{
Possible thermal effects to the motion of the $\tilde{N}_1$ are
discussed in Ref.~\cite{Hamaguchi:2001gw} which shows that those
effects are irrelevant as long as $M_1\gsim T_d$.
} 
As we see below, the flat direction $\phi$ is still stabilized at the
origin by the thermal mass term in the course of the $\tilde{N}_1$
oscillation. 

If $T_R>T_{\rm osci}$ (see Eq.~(\ref{eq:oscillate})), the $\tilde{N}_1$
starts to oscillate during the radiation dominated era, and hence
$|\tilde{N}_1|$ and $T$ decrease with $a(t)^{-3/2}$ and $a(t)^{-1}$,
respectively. 
Here, $a(t)$ denotes the scale factor of the universe.
To discuss the stability of $\phi$, it is convenient to define the 
temperature $T^{\rm back}\propto a(t)^{-1}$ during the $\tilde{N}_1$ 
domination, which corresponds to the temperature without the
$\tilde{N}_1$ decay.\footnote{ 
The actual background temperature is much higher than the temperature
$T^{\rm back}$, since the decay of the $\tilde{N}_1$ reheats up the
radiation. 
Thus, the condition in Eq.~(\ref{eq:stability1}) is a sufficient one to
stabilize the $LH_u$ flat direction by the thermal effects.
} 
Since $(T^{\rm back})^2$ decreases faster than $M_1|\tilde{N}_1|$,
$\phi=0$ is a stable point until the decay time of the $\tilde{N}_1$, if
the condition,   
\begin{eqnarray}
\label{eq:stability1}
 2 h M_1 |\tilde{N}_1^{\rm decay}| \ll \alpha^2 (T_d^{\rm back})^2,
\end{eqnarray}
is satisfied at the $\tilde{N}_1$ decay time.
Here, $T_d^{\rm back}$ is a background temperature at the $\tilde{N}_1$
decay time, which is given by  
\begin{eqnarray}
\label{eq:wouldbeTd}
 T_d^{\rm back}=T_{\rm dom}\bigg(\frac{a(t_{\rm dom})}{a(t_{\rm decay})}\bigg)
 = T_{\rm dom}\bigg(\frac{H_d}{H_{\rm dom}}\bigg)^{2/3}
 = T_{\rm dom}\bigg(\frac{T_d}{T_{\rm dom}}\bigg)^{4/3}
 = T_{\rm d}\bigg(\frac{T_d}{T_{\rm dom}}\bigg)^{1/3},
\end{eqnarray}
where $t_{\rm decay}$ denotes the decay time of the $\tilde{N}_1$, and
$H_{d,\,\rm dom}\propto T_{d,\,\rm dom}^2/M_{\rm pl}$ the Hubble
parameters at the decay time of the $\tilde{N}_1$ and at the beginning of
the $\tilde{N}_1$ domination, respectively.  
From the energy conservation at the decay time of the $\tilde{N}_1$, we
find that the amplitude  $|\tilde{N}_1^{\rm decay}|$ satisfies
\begin{eqnarray}
 \label{eq:decayamp}
  M_1^2 |\tilde{N}_1^{\rm decay}|^2 = \frac{\pi^2}{30} g_*(T_d)T_d^4.
\end{eqnarray}
Thus, the condition Eq.~(\ref{eq:stability1}) can be written as 
\begin{eqnarray}
 \label{eq:stab}
  \frac{\alpha^2}{2}\bigg(\frac{30}{\pi^2 g_*}\bigg)^{1/2}
  \bigg(\frac{T_d}{T_{\rm dom}}\bigg)^{2/3}
  \gg h.
\end{eqnarray}
By using the definition of $T_d$ in Eq.~(\ref{eq:energyconsv}) and 
$T_{\rm dom}< (90/\pi^2g_*)^{1/4}\sqrt{M_1 M_{\rm pl}}$, we obtain the  
sufficient condition for the $\phi$ stabilization as
\begin{eqnarray}
 \label{eq:stabcondition1}
  \frac{1}{4\pi}\bigg(\frac{\alpha^2}{2}\bigg)^3
  \bigg(\frac{30}{\pi^2 g_*}\bigg)^{3/2}
  \gg h. 
\end{eqnarray}
This is satisfied when $h$ satisfies the condition
Eq.~(\ref{eq:washout}).
Therefore, the flat direction $\phi$ remains at its origin if
$T_R>T_{\rm osci}$.   

On the other hand, if $T_R<T_{\rm osci}$, the $\tilde{N}_1$ starts to
oscillate before the completion of the reheating of inflation and the
situation becomes rather complex. 
Since $T$ decreases with $a(t)^{-3/8}$ during the inflaton dominated
era~\cite{Kolb:vq}, we should also require  
\begin{eqnarray}
 \label{eq:stability2}
  2 h M_1 |\tilde{N}_1^{\rm init}| \ll \alpha^2 T^2,
\end{eqnarray}
at the beginning of the $\tilde{N}_1$ oscillation for the stability of
$\phi$.
The temperature of the background radiation at the beginning of the
$\tilde{N}_1$ oscillation is estimated as
\begin{eqnarray}
 \label{eq:Tinit}
  T = T_R \bigg(\frac{a(t_R)}{a(t_{\rm osci})}\bigg)^{3/8}
  =T_R \bigg(\frac{H_{\rm osci}}{H_R}\bigg)^{1/4}
  \simeq (M_1 M_{\rm pl})^{1/4}T_R^{1/2},
\end{eqnarray}
where $t_R$ and $t_{\rm osci}$ denote the cosmic times of the end of the 
reheating and the beginning of the $\tilde{N}_1$ oscillation,
respectively, and $H_R$ and $H_{\rm osci}$ the Hubble parameters at 
those times. 
Here, we have used $H\propto a^{-3/2}$ in the inflaton dominated
era, $H_{\rm osci}\simeq M_1$, and $H_R =
(\pi^2g_*/90)^{1/4}T_R^2/M_{\rm pl}$.  
Thus, for $T_R<T_{\rm osci}$, we should also require in addition to
Eq.~(\ref{eq:stabcondition1})   
\begin{eqnarray}
\label{eq:stabcondition2}
 T_R \gg h \frac{2}{\alpha^2}\sqrt{M_1 M_{\rm pl}}
\bigg(\frac{|\tilde{N}_1^{\rm init}|}{M_{\rm pl}}\bigg) 
\simeq \frac{2\sqrt{4\pi}}{\alpha^2}T_d \bigg(\frac{|\tilde{N}_1^{\rm init}|}{M_{\rm pl}}\bigg),  
\end{eqnarray}
which is naturally satisfied in the $\tilde{N}_1$ dominated scenario.
Thus, we find that the flat direction $\phi$ remains also at the origin
for $T_R<T_{\rm osci}$.\footnote{ 
If the background temperature $T$ in Eq.~(\ref{eq:Tinit}) is larger than 
the inflaton mass $M_{\rm inf}$, the actual background temperature at
the beginning of the $\tilde{N}_1$ oscillation is $T\simeq M_{\rm inf}$
\cite{Kolb:2003ke}. 
Then, the condition in E.~(\ref{eq:stabcondition2}) is modified to
\begin{eqnarray}
 \alpha^2 M_{\rm inf}^2\gg h M_1 M_{\rm pl},
\end{eqnarray}
where $M_{\rm inf}$ is the inflaton mass.
}

Finally, we give a summary of the conditions which we should require to
the right-handed neutrino sector.
\begin{itemize}
 \item $M_1\ll H_{\rm inf}$; for a large initial amplitude of the
       $\tilde{N}_1$. 
 \item $T_R \gg T_d \,\, (3M_{\rm pl}^2/|N_1^{\rm init}|^2)$, for the
       $\tilde{N}_1$ domination. 
 \item $T_d < M_1$, to avoid washout effect of the lepton asymmetry.
 \item $v\gg H_{\rm inf}$, to fix $\tilde{X}$ and $\tilde{S}$ as in
       Eq.~(\ref{eq:minima}) during the inflation. 
 \item $f|\tilde{N}_1^{\rm init}|\ll v$, not to destabilize the
       $\tilde{S} \simeq v$.
 \end{itemize}
These conditions are easily satisfied, for example,
\begin{eqnarray}
\label{eq:parameters}
 M_1\simeq 10^{9 - 10} \,\,{\rm GeV},\,\,\,\,
  v \simeq 10^{15} \,\, {\rm GeV},\,\,\,\,   
  f \simeq 10^{-5},\,\,\,\,   
  h \simeq 10^{-6}.\,\,\,\,   
\end{eqnarray}

\section*{Acknowledgements}
The authors wish to thank K.Hamaguchi for a useful discussion on the
stabilization of the $LH_u$ flat direction.
This work is partially supported by Grand-in-Aid Scientific Research (s)
14102004.

\end{document}